\documentclass{aastex}
\usepackage{spr-astr-addons}
\usepackage{graphicx}
\shorttitle{Constraining annihilating dark matter by x-ray data}
\begin{document}
\title{Constraining annihilating dark matter by x-ray data}
\author{Man Ho Chan}
\affil{Department of Science and Environmental Studies, The Education University of Hong Kong \\
Tai Po, New Territories, Hong Kong, China}

\begin{abstract}
In the past decade, gamma-ray observations and radio observations put strong constraints on the parameters of dark matter annihilation. In this article, we suggest another robust way to constrain the parameters of dark matter annihilation. We expect that the electrons and positrons produced from dark matter annihilation would scatter with the cosmic microwave background photons and boost the photon energy to $\sim$ keV order. Based on the x-ray data from the Draco dwarf galaxy, the new constraints for some of the annihilation channels are generally tighter than the constraints obtained from 6 years of Fermi Large Area Telescope (Fermi-LAT) gamma-ray observations of the Milky Way dwarf spheroidal satellite galaxies. The lower limits of dark matter mass annihilating via $e^+e^-$, $\mu^+\mu^-$, $\tau^+\tau^-$, $gg$, $u\bar{u}$ and $b\bar{b}$ channels are 40 GeV, 28 GeV, 30 GeV, 57 GeV, 58 GeV and 66 GeV respectively with the canonical thermal relic cross section. This method is particularly useful to constrain dark matter annihilating via  $e^+e^-$, $\mu^+\mu^-$, $gg$, $u\bar{u}$ and $b\bar{b}$ channels.
\end{abstract}


\section{Introduction}
In the past few years, gamma-ray observations give some stringent constraints for annihilating dark matter. For example, recent studies of the Milky Way dwarf spheroidal satellite (MW dSphs) galaxies constrained the annihilation cross section lie below the canonical thermal relic cross section for dark matter of mass $m \sim 10-100$ GeV \citep{Ackermann,Li}. These results challenge the dark matter interpretation of the GeV gamma-ray excess observed in the Galactic center \citep{Daylan,Abazajian,Calore2,Calore,Abazajian2}. Generally speaking, this is a robust way to constrain dark matter annihilation because it involves direct detection of gamma rays. However, since some of the annihilation channels mainly give electrons and positrons (such as $e^+e^-$ and $\mu^+\mu^-$) \citep{Cirelli}, the gamma-ray constraints for these channels are not very stringent. The lower limits from the data of MW dSphs galaxies only give $m \ge 10$ GeV for $e^+e^-$ channel and $m \ge 5$ GeV for $\mu^+\mu^-$ channel \citep{Ackermann}.

Besides gamma-ray observation, radio observation is another useful way to constrain dark matter annihilation. The high-energy electron-positron pairs produced from dark matter annihilation would emit strong synchrotron radiation when there is a strong magnetic field \citep{Cirelli2}. For example, by using the radio observational data obtained in \citep{Davies} (at 408 MHz from the inner 4 arcsecond cone around Sgr A*), a very strong constraint on the dark matter annihilation cross section can be obtained \citep{Regis,Bertone}. The annihilation cross section for $m \sim 50$ GeV can be constrained to $<\sigma v> \le 10^{-27}$ cm$^3$ s$^{-1}$ \citep{Bertone,Cholis}. Furthermore, by using the radio data from \citet{Egorov,Giebubel}, the lower limits of dark matter mass can reach $\sim 100$ GeV for many annihilation channels \citep{Egorov,Chan}. However, this method is sensitively dependent on the magnetic field strength profile, which usually has large uncertainties \citep{Egorov,Giebubel2}. 

In this article, we suggest a third robust way to constrain dark matter annihilation. Since the electron-positron pairs produced would scatter with the cosmic microwave background (CMB) photons via inverse Compton scattering (ICS), these electrons and positrons can boost the photon energy up to keV order \citep{Colafrancesco,Beck}. By using the x-ray data from MW dSphs galaxies, we can get a tight constraint for the annihilation cross section for each annihilation channel. Since the magnetic field strength for MW dSphs is $0.1-1$ $\mu$G \citep{Colafrancesco,Beck}, ICS is the major cooling mechanism for the high-energy electron and positron pairs. Therefore, the only free parameters in this method are dark matter mass and the annihilation cross section. In the following analyses, we mainly focus on six different standard model annihilation channels ($e^+e^-, \mu^+\mu^-, \tau^+\tau^-$, $gg$, $u\bar{u}$ and $b\bar{b}$) for $m=10-100$ GeV.

\section{ICS of the CMB photons in Draco}
Generally speaking, an electron with a Lorentz factor $\gamma$ can increase photon energy from $E_0$ to $\sim \gamma^2E_0$ via ICS. Since the original energy of the CMB photons is about $6 \times 10^{-4}$ eV, a 1 GeV high-energy electron ($\gamma \sim 2000$) can boost the energy of a CMB photon to about $\sim 1$ keV (the full spectrum of the ICS photons can be ranging from 0.1-20 keV). Therefore, these high-energy photons can be detected by x-ray observations. In particular, dwarf galaxies are the best objects for this analysis because they are dark matter-dominated and their magnetic field strengths are weak $B \sim 0.1-1$ $\mu$G \citep{Colafrancesco,Beck}. Therefore, the uncertainties in the magnetic field strength can be minimized. In the following, we would use the x-ray data of the Draco dwarf galaxy to perform the analysis. Also, for dark matter mass $m \ge 100$ GeV, the resulting ICS photons would be mainly MeV photons. Since the sensitivity of MeV photon detection is not good enough to constrain dark matter, we will just focus on $m=1-100$ GeV.

The energy spectrum $dN'/dE'$ of the electrons produced from dark matter annihilation via different channels can be obtained in \citep{Cirelli}. The number of CMB photons scattered per second from original frequency $\nu_0$ to new frequency $\nu$ via ICS is given by
\begin{equation}
I(\nu)=\frac{3 \sigma_Tc}{16 \gamma^4} \frac{n(\nu_0)\nu}{\nu_0^2} \left[2 \nu \ln \left(\frac{\nu}{4\gamma^2\nu_0} \right)+\nu+4\gamma^2 \nu_0- \frac{\nu^2}{2\gamma^2 \nu_0} \right],
\end{equation}
where $\sigma_T$ is the Thomson cross section and $n(\nu_0)=170x^2/(e^x-1)$ cm$^{-3}$ is the number density of the CMB photons with frequency $\nu_0$, where $x=h\nu_0/kT_{\rm CMB}$. However, \citet{Colafrancesco} show that the diffusion of electrons in the Draco dwarf galaxy might be important so that the resultant ICS signal is significantly suppressed. In fact, the diffusion coefficient for dwarf galaxies is highly uncertain. Generally speaking, there are three different types of diffusion models to describe the diffusion processes of high-energy electrons, namely the Kolmogorov model, the Kraichnan model and the Boehm model (the random walk model) \citep{Regis}. The range of the diffusion coefficient ($10^{26}-10^{28}$ cm$^2$ s$^{-1}$) assumed in \citet{Colafrancesco} is based on the Kolmogorov model, which is commonly used to model our Milky Way. However, the effect of turbulence and irregularities of magnetic field in the Draco dwarf galaxy is quite different from that in our Milky Way. As shown in \citet{Strong,Regis}, the diffusion coefficient is given by $D=(1/3)r_gv_e(\delta B/B)^{-2}$, where $r_g$ is the gyroradius of electrons, $v_e$ is the speed of electrons, $\delta B$ is the magnetic field irregularities and $B$ is the mean magnetic field strength. For our Milky Way, $\delta B/B$ is very small ($\delta B \sim \mu$G and $B \ge 10~ \mu$G) so that the electrons do not `feel' the fine structure in the magnetic field but move in orbits determined by the mean magnetic field which is much greater in magnitude than the fluctuating component \citep{Longair}. This would greatly enhance the diffusion process. However, for the Draco dwarf galaxies, it does not contain much interstellar medium so that the effect of turbulence is not large. Even if there exists some irregularities $\delta B \sim B \sim \mu$G, the term $\delta B/B$ is of the order 1. As a result, the diffusion coefficient would be close to $D \sim (1/3)r_gv_e$, which is best described by the Boehm model, the random walk model without any turbulence \citep{Regis}.

Here, let's apply the Boehm model to estimate the diffusion length of the high-energy electrons. For the magnetic field strength $B \sim 1$ $\mu$G, the Larmor radius is $r_L \sim 10^{12}$ cm. Since the cooling time of a 1 GeV electron is about $t_c \sim 10^{16}$ s, the stopping distance of a 1 GeV electron is $d_s \sim \sqrt{r_Lct_c} \sim 10^{-3}$ kpc \citep{Boehm}, which is much shorter than the core radius of the Draco dwarf galaxy ($\sim 10^{-1}$ kpc). That means that the diffusion process is not very important. Based on this assumption, the electron number density energy distribution function can be simply given by \citep{Storm}
\begin{equation}
\frac{dn_e}{dE}(\tilde{E})=\frac{<\sigma v> \rho^2}{2m^2b(\tilde{E})} \int_{\tilde{E}}^{m} \frac{dN'}{dE'}dE',
\end{equation}
where $\rho$ is the dark matter density profile, $<\sigma v>$ is the annihilation cross section and $b(\tilde{E})$ is the cooling rate, which is given by \citep{Colafrancesco2}
\begin{equation}
b(\tilde{E})=\left[0.25\tilde{E}^2+0.0254 \left( \frac{B}{\rm 1~\mu G} \right)^2 \tilde{E}^2 \right] \times 10^{-16}~{\rm GeV/s},
\end{equation}
with $\tilde{E}$ in GeV. Since the thermal electron number density is low $n \sim 10^{-6}$ cm$^{-3}$ \citep{Colafrancesco}, we neglect the Bremsstrahlung and Coulomb cooling. If we assume $B=1$ $\mu$G, the ICS would dominate the cooling process and the resulting cooling rate is $b(\tilde{E}) \approx 0.275 \times 10^{-16}\tilde{E}^2$ GeV/s. 

The total x-ray energy flux in the energy band $E_1$ to $E_2$ is given by
\begin{equation}
\Phi=2\times \frac{<\sigma v>J}{8\pi m^2} \int_{E_1}^{E_2}d(h\nu) \int_{m_e}^{m}\frac{Y(\tilde{E})}{b(\tilde{E})}d\tilde{E} \int_0^{\infty}I(\nu)dx,
\end{equation}
where
\begin{equation}
J=\int_{\Delta \Omega}d\Omega \int_{\rm los} \rho^2 ds
\end{equation}
is called the J-factor and
\begin{equation}
Y(\tilde{E})=\int_{\tilde{E}}^m \frac{dN'}{dE'}dE'.
\end{equation}
The dark matter density profile $\rho$ for Draco can be modeled by
\begin{equation}
\rho=\rho_0\left[1+\left(\frac{r}{r_0} \right)^{1.5} \right]^{-2.25},
\end{equation}
where $\rho_0=0.65 \times 10^9M_{\odot}$ kpc$^{-3}$ and $r_0=0.28$ kpc \citep{Gilmore,Riemer}. By including the substructure contribution (the boost factor is 3.43 \citep{Beck}), the resulting J-factor of the core region of Draco is $\log(J/\rm GeV^2~cm^{-5})=19.1$. Recent empirical fits of kinematic data of Draco give another set of parameters, which has a smaller core density but a larger core radius \citep{Burkert}. The density profile assumed is the King profile \citep{King}. The resultant J-factor by using this set of parameters gives $\log(J/\rm GeV^2~cm^{-5}) \approx 19.2$. Since the dark matter profile in Eq.~(7) gives a smaller J-factor, we will use it to calculate the predicted flux as it can obtain the most conservative bounds for annihilating dark matter. Fig.~1 shows the flux spectrum of the resulting photons. Most of the photons lie between $0.1-1$ keV, which can be seen by x-ray observations. 

The x-ray energy flux ($0.1-2.4$ keV) of the central core part of Draco is $\Phi<1.7 \times 10^{-14}$ erg cm$^{-2}$ s$^{-1}$ \citep{Zang,Colafrancesco}. If we assume that all the x-ray luminosity originates from the ICS of the CMB photons and use the canonical thermal relic cross section \citep{Steigman}, we can put an upper limit for each of the annihilation channels. Fig.~2 shows the total x-ray energy fluxes for the six annihilation channels. We can see that all the annihilation channels are ruled out for $m \le 28$ GeV. In particular, the most popular model for the dark matter interpretation of the GeV excess ($m=40-60$ GeV via $b\bar{b}$ \citep{Calore,Abazajian2}) is also in considerable tension (see table 1). 

If we release the annihilation cross section to be a free parameter, we can get the constraints of the cross section against dark matter mass for different annihilation channels. Fig.~3 shows that our method gets tighter constraints than the results from Fermi-LAT for the $e^+e^-$ and $\mu^+\mu^-$ channels \citep{Ackermann}. It also works well for the $gg$, $u\bar{u}$ and $b\bar{b}$ channels. However, the constraint for the $\tau^+\tau^-$ channel is not very stringent because only less than 20\% of the energy from dark matter annihilating via $\tau^+\tau^-$ is shared to the electron and positron pairs \citep{Cirelli}. Overall speaking, this method can be regarded as another robust test to constrain annihilating dark matter parameters.

Note that we have only included the CMB energy density in our calculations. In fact, there are other radiation fields in the infra-red and visible light bands which can contribute to the x-ray flux via ICS. Therefore, our results are conservative bounds because the upper bound of the observed x-ray energy flux may include the effect of those non-CMB photons. Nevertheless, the contribution of other radiation fields is small because most of the resulting photons via ICS are in MeV or above bands. Also, the overall intensity due to other radiation field is much less than the effect of CMB photons (see \citep{Colafrancesco}).

\begin{table}
\caption{Lower limits of dark matter mass for different annihilation channels. Here, we assume $<\sigma v>=2.2 \times 10^{-26}$ cm$^3$ s$^{-1}$.}
 \label{table1}
 \begin{tabular}{@{}lc}
  \hline
   & Lower limit (GeV) \\
  \hline
  $e^+e^-$ & 40 \\
  $\mu^+\mu^-$ & 28 \\
  $\tau^+\tau^-$ & 30 \\
  $gg$ & 57 \\
  $u\bar{u}$ & 58 \\
  $b\bar{b}$ & 66 \\
  \hline
 \end{tabular}
\end{table}

\begin{figure}
\vskip 10mm
 \includegraphics[width=82mm]{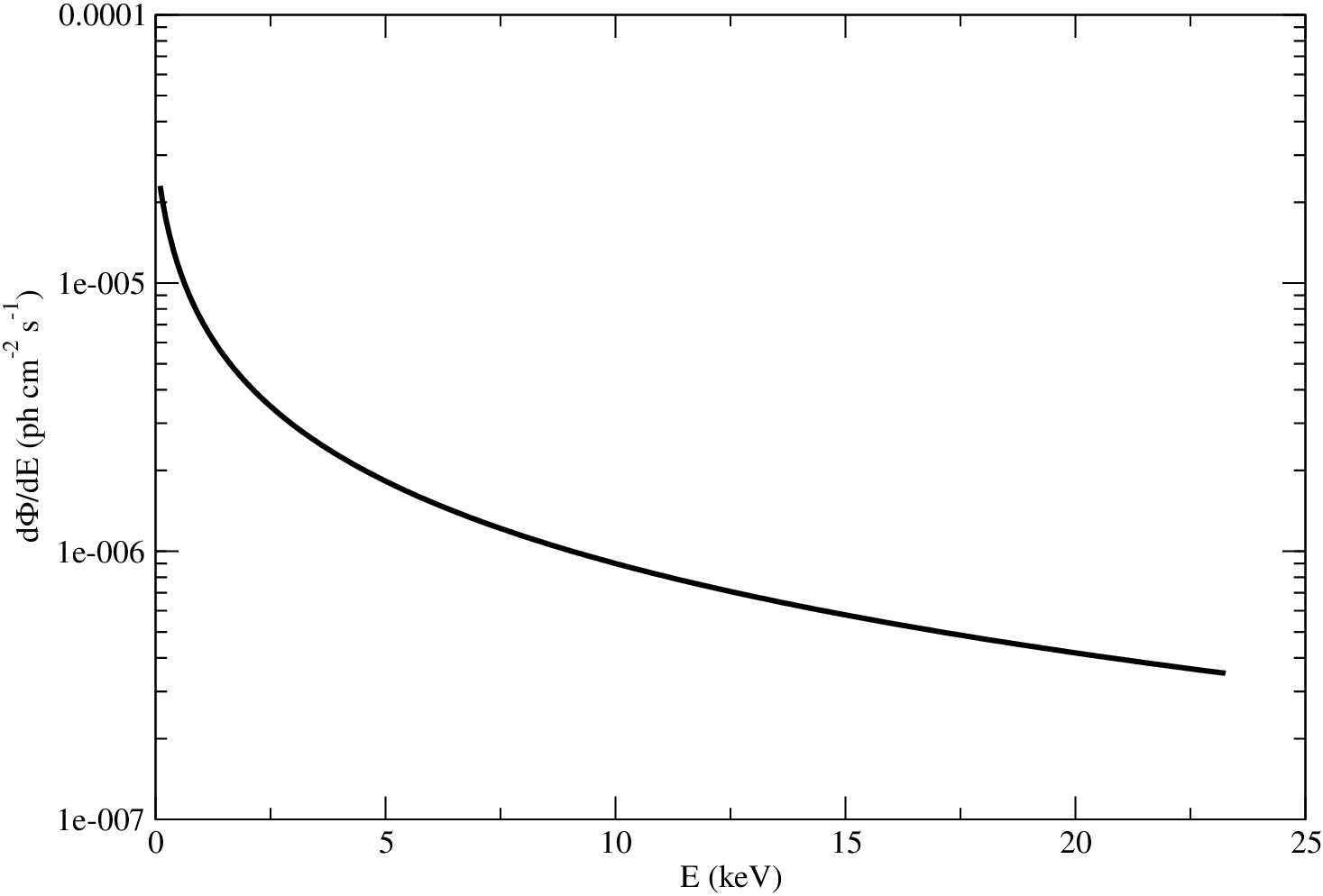}
 \caption{The resulting photon spectrum. Here, we assume $<\sigma v>=2.2 \times 10^{-26}$ cm$^3$ s$^{-1}$.}
\vskip 10mm
\end{figure}

\begin{figure}
\vskip 10mm
 \includegraphics[width=82mm]{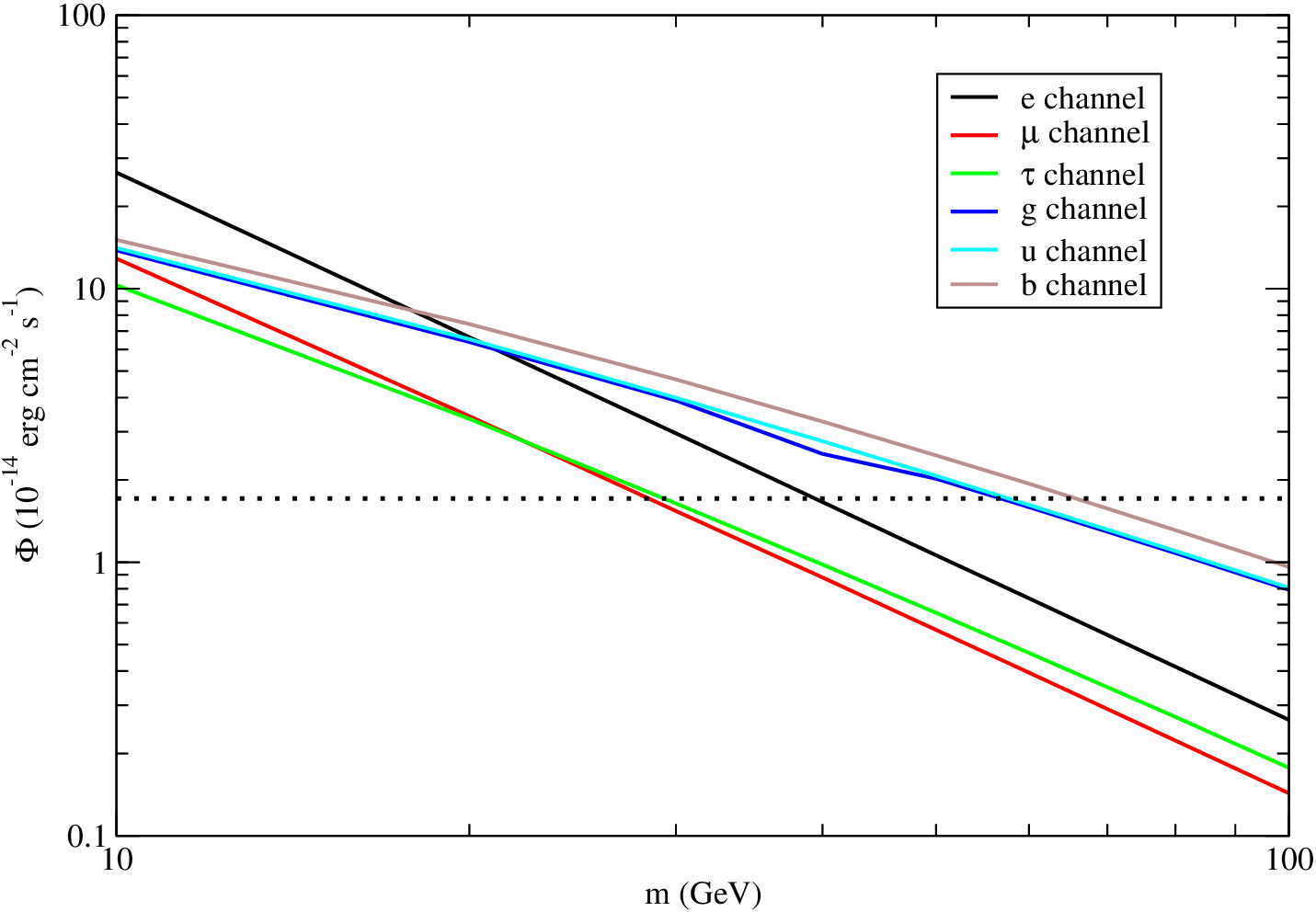}
 \caption{The predicted x-ray flux for the six annihilation channels (due to ICS of the CMB photons). The dotted line represents the upper limit of the observed x-ray flux for the Draco dwarf galaxy. Here, we assume $<\sigma v>=2.2 \times 10^{-26}$ cm$^3$ s$^{-1}$.}
\vskip 10mm
\end{figure}

\begin{figure}
\vskip 10mm
 \includegraphics[width=82mm]{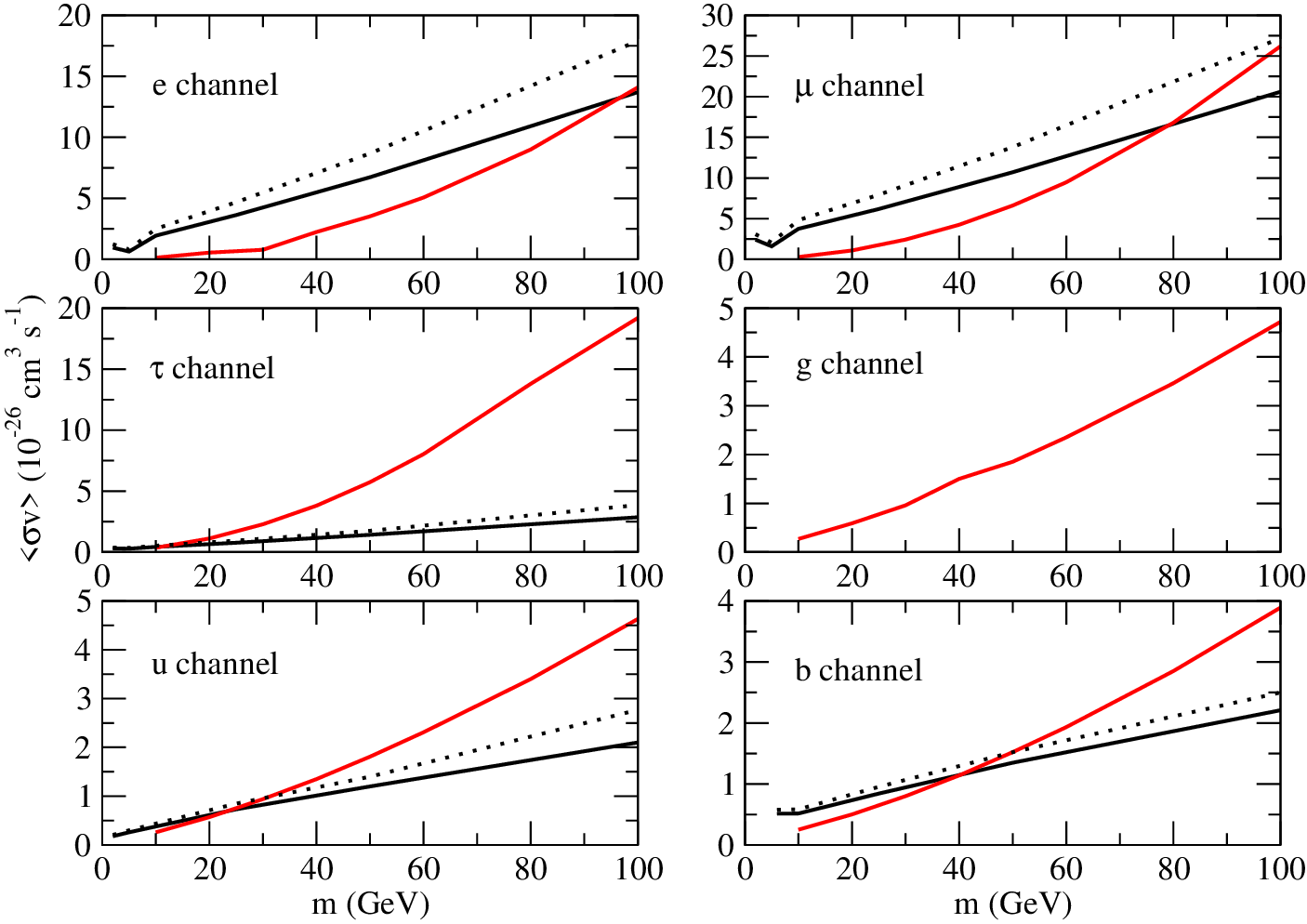}
 \caption{The upper limits of the annihilation cross sections for the six annihilation channels. The red solid lines represent our results. The black solid lines represent the gamma-ray observations of MW dSphs galaxies with the NFW density profile (black dotted lines: with J-factor uncertainties) \citep{Ackermann}. Note that the Fermi-LAT gamma-ray observations in \citet{Ackermann} did not perform the analysis for the $gg$ channel.}
\vskip 10mm
\end{figure}

\section{Discussion}
In this article, we propose another new method to constrain the parameters of annihilating dark matter. By calculating the resulting photon energy spectrum due to ICS of the CMB photons, we can determine the upper limits of the annihilation cross sections by x-ray data of dwarf galaxies. In this analysis, we use the Draco dwarf galaxy because it has a relatively small uncertainty in the J-factor estimation \citep{Bonnivard} and it has a better x-ray constraint. This method can give more stringent constraints on annihilation cross sections for some of the annihilation channels compared with the gamma-ray observations, especially for the $e^+e^-$ channel, $\mu^+\mu^-$ channel, $u\bar{u}$ channel for $m \le 20$ GeV and $b\bar{b}$ channel for $m \le 40$ GeV. Since the magnetic field strength is small $B \sim 1$ $\mu$G and the thermal electron number density is low $n \sim 10^{-6}$ cm$^{-3}$ for dwarf galaxies, the cooling rate is dominated by the ICS cooling. Hence, this method is particularly good to apply in dwarf galaxies. Besides, since the CMB photon density is nearly uniform for all dwarf galaxies, we do not have any free parameter except the dark matter mass and the annihilation cross section.

Generally speaking, this method is better than the analysis of radio data. Although the constraints from radio data are usually more stringent \citep{Egorov,Chan}, the systematic uncertainties due to the magnetic field strength profile are large. Nevertheless, in our method, by using the data from dwarf galaxies with weak magnetic field strength $B \le 1$ $\mu$G, we can obtain good constraints with a small systematic uncertainty. The only systematic uncertainty comes from the J-factor, which must exist in this kind of analyses, including the gamma-ray observations \citep{Ackermann}.

By comparing our results with the recent result based on cosmic reionization \citep{Liu}, we get similar upper bound for the $e^+e^-$ channel. The upper bound of dark matter mass for the $e^+e^-$ channel for the canonical thermal relic cross section is $m \le 30$ GeV for s-wave annihilation \citep{Liu}. Our result ($m \le 40$ GeV) is a bit tighter than this bound.

Based on our analyses, most of the current popular dark matter models are in considerable tension with our results. For example, \citet{Calore} show that dark matter annihilating via $b\bar{b}$ with $m=48.7^{+6.4}_{-5.2}$ GeV and $<\sigma v>=1.75^{+0.28}_{-0.26} \times 10^{-26}$ cm$^3$ s$^{-1}$ or $gg$ with $m=57.5^{+7.5}_{-6.5}$ GeV and $<\sigma v>=2.16^{+0.35}_{-0.32}\times 10^{-26}$ cm$^3$ s$^{-1}$ can give good explanation for the Galactic center GeV gamma-ray excess. However, the allowed ranges become narrower if our results are included, especially for the $b\bar{b}$ channel. 

To conclude, this method gives a new research direction in constraining dark matter annihilation. Future x-ray observations for dwarf galaxies can definitely give a better constraint for annihilating dark matter and verify the existing dark matter models.

\section{acknowledgements}
This work is supported by a grant from The Education University of Hong Kong (Project No.:RG4/2016-2017R).

\bibliographystyle{spr-mp-nameyear-cnd}
\bibliography{biblio-u1}

\clearpage

\end{document}